\documentclass[conference,letterpaper]{IEEEtran}

\usepackage{subfigure}
\usepackage{moreverb}
\usepackage{epsfig}
\usepackage{amsmath,amssymb,amsthm,mathrsfs,amsfonts,dsfont}
\usepackage{adjustbox,lipsum}
\usepackage{algorithm,algorithmic}
\usepackage{amsfonts}
\usepackage{epsfig}
\usepackage{amssymb}
\usepackage{amsmath}
\usepackage{amsthm}
\usepackage{subfigure}
\usepackage{multirow}
\usepackage{rotating}
\usepackage{graphicx}
\usepackage{tabularx}
\usepackage{array}
\usepackage{anyfontsize}
\usepackage{color,soul}
\usepackage{graphicx,dblfloatfix}
\usepackage{epstopdf}
\usepackage{blindtext}
\usepackage{amsmath}
\usepackage{amsthm,amssymb,amsmath,bm}
\usepackage{subfigure}
\usepackage{amsfonts}
\usepackage{epsfig}
\usepackage{amssymb}
\usepackage{amsmath}
\usepackage{cite}
\usepackage{graphicx}
\usepackage{fancyhdr}
\usepackage{subfigure}
\usepackage[subfigure]{tocloft}
\usepackage[font={small}]{caption}
\usepackage{subfigure}
\usepackage{tabularx}
\usepackage{tcolorbox}
\usepackage{cite}

\begin{document}

\title{ Power Minimization  in  Wireless Sensor Networks  With
	Constrained AoI Using  Stochastic Optimization
\author{  \IEEEauthorblockN{Mohammad Moltafet and  Markus Leinonen  }
	\IEEEauthorblockA{Centre for Wireless Communications -- Radio Technologies\\
		University of Oulu, Finland\\
		e-mail: $\{$mohammad.moltafet, markus.leinonen$\}$@oulu.fi}	
	\and
	\IEEEauthorblockN{  Marian Codreanu and Nikolaos Pappas}
	\IEEEauthorblockA{Department of Science and Technology\\
		Link\"{o}ping University, Sweden\\
		e-mail: $\{$marian.codreanu, nikolaos.pappas$\}$@liu.se}
} 
}
\maketitle
\begin{abstract} 
	In this paper, we consider a system where multiple low-power sensors communicate timely information about a random process to a sink. The sensors share orthogonal subchannels to transmit such information in the form of status update packets. Freshness of the sensors' information at the sink is characterized by the Age of Information (AoI), and the sensors can control the sampling policy by deciding whether to take a sample or not. We formulate an optimization problem to minimize the time average total transmit power of sensors by jointly optimizing the sampling action of each sensor, the transmit power allocation, and the subchannel assignment under the constraints on the maximum time average AoI and maximum power of each sensor. To solve the optimization problem, we use the Lyapunov drift-plus-penalty method. Numerical results show the performance of the proposed algorithm versus the different parameters of the system.
	
\end{abstract}	
%
%
\allowdisplaybreaks

\allowdisplaybreaks

\section{Introduction}

Freshness of the status information of various physical processes collected by multiple sensors is a key performance enabler in many applications of wireless sensor networks (WSNs) \cite{8187436,8469047,8901143}, e.g., surveillance in smart home systems and drone control. The Age of Information (AoI) was introduced as a destination centric metric that characterizes the freshness in status update systems \cite{6195689,6310931,5984917}.
A status update packet  of each sensor contains a time stamp representing  the time when the sample was generated and the measured value of the monitored process.  If at a time instant $t$, the most recently received status update packet
contains the time stamp $U(t)$, the AoI is defined
as  $\Delta(t)=t-U(t)$. In other words,  the AoI of each sensor  is the time elapsed since the last received status update packet was generated at the sensor. The average AoI is the most commonly used  metric to evaluate the AoI \cite{8469047,6195689,6310931,5984917,8486307,8006703,chen2019optimal,8606155,8901143,8123937,moltafet2019age}.

The authors of \cite{8486307} considered a WSN in which sensors share one  unreliable subchannel  in each slot. 
 They minimized the  expected weighted sum AoI of the network
 by determining the transmission scheduling policy. The authors of \cite{8006703} considered  an energy harvesting sensor and derived  the optimal threshold in terms of  remaining energy to trigger a new sample. The authors of  \cite{8123937} considered  an energy harvesting sensor and minimized the time average AoI by determining the optimal status update policy.  The authors of \cite{chen2019optimal} considered  two source nodes generating heterogeneous traffic with different power supplies  and  studied the peak-age-optimal  status update scheduling. The authors of \cite{8606155} considered a  wireless power
 transfer powered sensor network  and studied performance of the system  in terms of the average  AoI.

In this paper, we minimize  the time average total transmit power of sensors by jointly optimizing  the sampling action, the transmit power allocation, and the subchannel assignment in each slot under the constraints on the maximum time average AoI and maximum power of each
sensor. 
To solve the proposed optimization problem, we apply the Lyapunov drift-plus-penalty method. To the best of our knowledge, joint optimization of the transmit power allocation, subchannel assignment, and sampling action with constrained AoI  has not been studied earlier.
The most related work to this paper is  \cite{8486307}.  Differently from \cite{8486307}, besides the sampling action of each sensor, we  consider both transmit power allocation and subchannel assignment in each slot.  

\section{ System Model and Problem Formulation}\label{System Model and Problem Formulation}
\begin{figure}
	\centering
	\includegraphics[scale=.4]{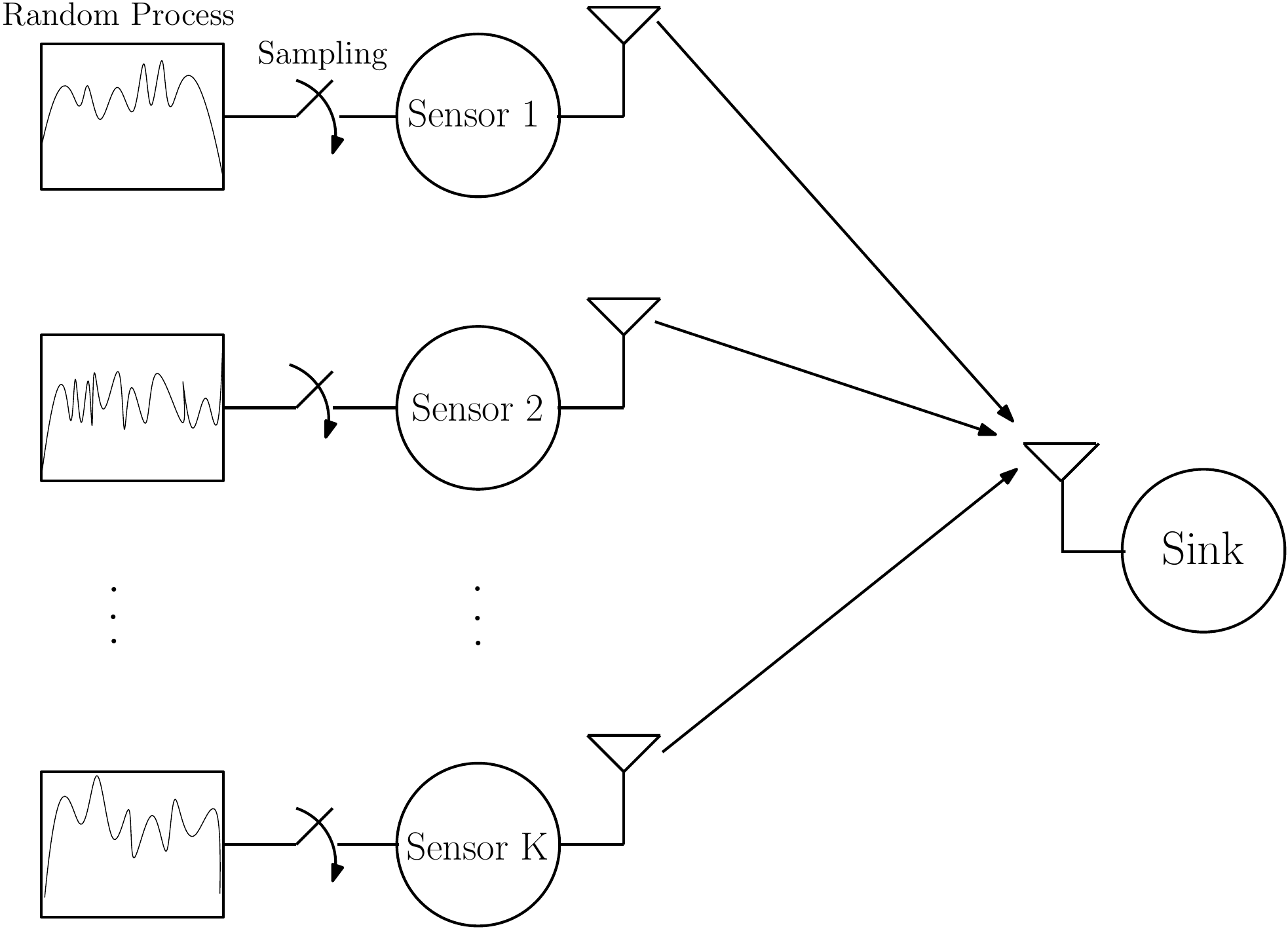}
	\caption{System model.}
	\vspace{-5mm}
	\label{model}
\end{figure}

We consider a WSN consisting of a set $\mathcal{K}$ of $K$ sensors and one sink,  as depicted  in Fig. \ref{model}. The sink is interested in time-sensitive information from the sensors which measure a physical phenomenon.  We assume  a slotted communication with normalized slots ${t\in\{1,2,\dots\}}$, where in each slot, the sensors share a set $\mathcal{N}$ of $N$ orthogonal subchannels  with bandwidth $W$ per subchannel. We consider that each sensor can control the sampling process by deciding whether to take a sample or not
at the beginning of each slot $t$. We assume that the perfect channel state information is available at the sink. 

Let $\rho_{k,n}(t)$ denote the subchannel assignment at time slot $t$ as $\rho_{k,n}(t)\in \{0,1\}, \forall k \in\mathcal{K}, n \in\mathcal{N}$, where   $\rho_{k,n}(t)=1$ indicates that subchannel $n$ is assigned to sensor $k$ at  time slot $t$, and  $\rho_{k,n}(t)=0$ otherwise. To ensure that at any given time slot $t$, each subchannel can be assigned to at most one sensor, the following constraint is used:
\begin{align}\label{mn001}
\textstyle\sum_{k\in\mathcal{K}}\rho_{k,n}(t)\le 1,  n\in \mathcal{N}, \forall t.
\end{align}

Let $p_{k,n}(t)$ denote the transmitted power of sensor $k$ over subchannel $n$ at slot $t$.
Then,  the signal-to-noise ratio  (SNR)  with respect to sensor $k$ over subchannel $n$ at slot $t$ is given by 
\begin{align}
\gamma_{k,n}(t)=\dfrac{p_{k,n}(t)|h_{k,n}(t)|^2}{WN_0},
\end{align}
where $h_{k,n}(t)$ is the channel coefficient from sensor $k$ to the sink over subchannel $n$ at slot $t$ and $N_0$ is the noise power spectral density. Accordingly, the achievable  rate for sensor $k$ over  subchannel $n$ in slot $ t $  is given by
\begin{align}
r_{k,n}(t)=W\log_2\left(1+\gamma_{k,n}(t)\right).
\end{align}
The achievable data rate of sensor $k$ at slot $t$ is equal 
to the summation of achievable data rates over all assigned subchannels at  slot $t$,   expressed  as 
$$
R_{k}(t)=\textstyle\sum_{n\in\mathcal{N}}\rho_{k,n}(t)r_{k,n}(t).
$$

   Let  $b_k(t)$ denote the sampling action of  sensor $k$ at time slot $t$  as $b_k(t)\in \{0,1\}, \forall k \in\mathcal{K}$, where   $b_k(t)=1$ indicates that sensor  $k$  takes a sample at the beginning of time slot $t$, and  $b_k(t)=0$ otherwise. We assume that  sampling time (i.e.,  the time needed to take a sample) is negligible. 
We consider that sensor $k$ takes a sample at the beginning of slot $t$ only if there are enough resources  to transmit the sample during the same slot $t$. 
In other words, if sensor $k$ takes a sample at the beginning of slot $t$ (i.e., $b_k(t)=1$), the sample will be transmitted during the same slot $t$.   To this end, we use the following constraint:
\begin{align}
R_{k}(t)= \eta b_{k}(t),  k\in \mathcal{K}, \forall t,
\end{align}
 where  $\eta$ is the  size of each status update packet (bits).
This constraint ensures that when sensor $k$ takes a sample at the beginning of slot $t$ (i.e., $b_k(t)=1$), the achievable rate for sensor $k$ at slot $t$  is $R_{k}(t)= \eta$, which guarantees that the sample is transmitted during the slot.

Let $\delta_k(t)$ denote the AoI of the sensor $k$ at the beginning of slot $t$. 
 If sensor $k$ takes a sample at the beginning of slot $t$ (i.e., $b_k(t)=1$),  the AoI  at the beginning of slot $t+1$ drops to one, and
   otherwise (i.e., $b_k(t)=0$), the AoI  is incremented by one. Accordingly, the evolution of $\delta_k(t)$ is  characterized as 
\begin{align}\label{AoI1}
\delta_k(t+1)&=\begin{cases}
1&,\text{if} \,\,b_k(t)=1;\\
\delta_k(t)+1&,\text{otherwise}.
\end{cases}
\end{align}

\begin{figure}
	\centering
	\includegraphics[scale=.9]{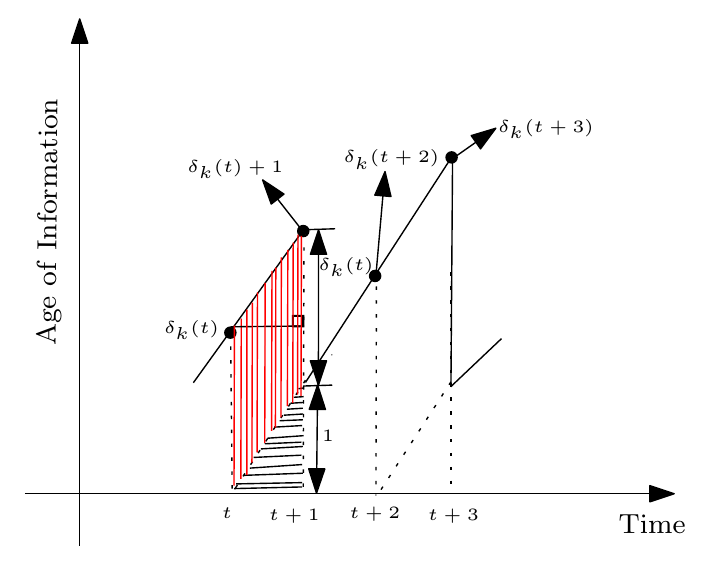}
	\caption{The evolution of the AoI of sensor $k$.}
	 		\vspace{-5mm}
	\label{AoI}
\end{figure}

The evolution of the AoI of  sensor $k$ is illustrated in Fig. \ref{AoI}.  The time average AoI of sensor $k$ is calculated as the  area under the AoI curve, normalized by the observation interval. As it can be seen, during slot $t$, the area under the AoI curve of sensor $k$ is calculated as a sum of the areas of a triangle and a parallelogram.  The area of the triangle is equal to $1/2$ and the area of the  parallelogram is equal to $\delta_k(t)$. {Therefore, the time average AoI of sensor $k$ is calculated as 
\begin{align}\label{mnbhg010}
\Delta_k&=
\dfrac{1}{2}+\lim_{t\to \infty}\dfrac{1}{t}\textstyle\sum_{\tau=1}^{t}\delta_k(\tau).
\end{align}}

{To make the calculations tractable, we use a commonly used approach that instead of the time average AoI in \eqref{mnbhg010}, 
we consider the time average of expectation of the AoI \cite{8486307,neely2010stochastic,stochast009om}, given as 
\begin{align}\label{mnbhg01}
\Delta_k&=
\dfrac{1}{2}+\lim_{t\to \infty}\dfrac{1}{t}\textstyle\sum_{\tau=1}^{t}{\mathbb{E}}[\delta_k(\tau)],
\end{align}
where the expectation is with respect to the  random wireless channel states and control actions made in reaction to the channel states\footnote{Through the paper, all expectations are taken with respect to the randomness of the wireless channel states and control actions made in reaction to the channel states.}.}  We consider that the initial value of the AoI of all sensors is   $\delta_k(1)=0, \,\,\forall k\in \mathcal{K}$. 
\subsection{Problem Formulation}
Our objective  is to minimize the time average total transmit power of sensors by optimizing  the sampling action, the transmit power allocation, and the subchannel assignment in each slot subject to  the maximum  time average AoI   and maximum power constraints for each sensor.
Thus, the optimization problem is formulated as follows
 \begin{subequations}\label{eqo1}
 	\begin{align}\label{eq8a}
 	&\text{minimize} \,\,\lim_{t\to \infty}\dfrac{1}{t}\textstyle\sum_{\tau=1}^{t}\textstyle\sum_{k\in \mathcal{K}}\textstyle\sum_{n\in\mathcal{N}}\mathbb{E}[p_{k,n}(\tau)]\\&\label{eq8o2} 
 	\text{subject\,\,to}\hspace{.37cm}
\textstyle\sum_{k\in\mathcal{K}}\rho_{k,n}(t)\le 1, \forall n\in \mathcal{N}, \forall t\\&\label{eqo3}
 	\hspace{1.8cm}\textstyle\sum_{n\in\mathcal{N}}p_{k,n}(t)\le P_k^{\text{max}},\,\,\, k\in \mathcal{K}, \forall t\\&\label{eqo5}
 	\hspace{1.8cm}\Delta_k\le \Delta^{\text{max}}_k,\,\,\, k\in \mathcal{K}
 	\\&\label{eq1o5}
 	\hspace{1.8cm}\textstyle\sum_{n\in\mathcal{N}}\rho_{k,n}(t)r_{k,n}(t)= \eta b_{k}(t),  k\in \mathcal{K}, \forall t
 	 \\&\label{eq1o50}
 	\hspace{1.8cm}\rho_{k,n}(t)\in\{0,1\},  k\in \mathcal{K}, n\in \mathcal{N}, \forall t
 	 \\&\label{eq1o51}
 	\hspace{1.8cm} b_{k}(t)\in\{0,1\},  k\in \mathcal{K},  \forall t,
 	\end{align}
 \end{subequations}
with variables $\{p_{k,n}(t),\rho_{k,n}(t)\}_{k \in\mathcal{K}, n \in\mathcal{N}}$ and $\{b_{k}(t)\}_{k \in\mathcal{K}}$ for all ${t\in\{1,2,\dots\}}$. The constraints of  problem \eqref{eqo1} are as follows.
The inequality \eqref{eq8o2} constrains  that each subchannel can be assigned to at most one sensor in each slot; the inequality 
 \eqref{eqo3}  constrains the power of each sensor with respect to the maximum budget $P_k^{\text{max}}$; the inequality  \eqref{eqo5} is the maximum acceptable time average  AoI constraint for each sensor; the equality  \eqref{eq1o5} 
 ensures that each sample is transmitted during one slot;  \eqref{eq1o50} and \eqref{eq1o51} represent the feasible values for the subchannel assignment and sampling policy variables, respectively. 
 
The proposed optimization problem is a mixed integer programming problem where the constraints and the objective function both contain time averages over the optimization variables. In the following section,  a dynamic control algorithm using the Lyapunov optimization approach is presented to solve optimization problem \eqref{eqo1}.

\section{Solution Algorithm}\label{Solution Algorithm}
We use the Lyapunov drift-plus-penalty method  introduced in \cite{neely2010stochastic} and \cite{stochast009om} to solve the optimization problem \eqref{eqo1}. According to the drift-plus-penalty method, the time average constraint \eqref{eqo5} is enforced by transforming the problem into a queue stability problem. In other words, for each time average inequality in constraint \eqref{eqo5} a virtual queue is associated in such a way that the stability of these virtual queues implies the feasibility of constraint \eqref{eqo5}.

To use the drift-plus penalty method, we rewrite constraint \eqref{eqo5} as follows
\begin{align}\label{consta1}
\lim_{t\to \infty}\dfrac{1}{t}\textstyle\sum_{\tau=1}^{t}\mathbb{E}[\delta_k(\tau)]\le \Delta^{\text{max}}_k-\dfrac{1}{2},\,\,\, k \in\mathcal{K}.
\end{align}
Let $\{Q_k(t)\}_{k\in \mathcal{K}}$ denote the virtual queues associated with constraint  \eqref{consta1}. Then, the virtual queues are updated at the beginning of each time slot as 
\begin{align}\label{consta2}
Q_k(t\!+\!1)\!=\!\max\!\left[Q_k(t)-\left(\Delta^{\text{max}}_k\!\!-\!\!\dfrac{1}{2}\right),0\right]\!\!+\!\delta_k(t\!+\!1), \forall k\!\in\! \mathcal{K}.
\end{align}
 Here, we use the notion of strong stability; the virtual queues are strongly stable if \cite[Ch. 2]{neely2010stochastic} 
\begin{align}\label{consta3}
\lim_{t\to \infty}\dfrac{1}{t}\textstyle\sum_{\tau=1}^{t}\mathbb{E}[Q_k(\tau)]<\infty, \forall k\in \mathcal{K}.
\end{align}
According to \eqref{consta3}, a queue is strongly stable if its time average backlog is finite. Next, we introduce the Lyapunov function and its drift  which are needed to define the queue stability problem. 
%

 Let $\bold{S}(t)=\{Q_k(t),\delta_k(t)\}_{k\in \mathcal{K}}$ denote the network state at the beginning of slot $t$,  and $\bold{Q}(t)$ denote a vector containing all the virtual queues, i.e., ${\bold{Q}(t)=[Q_1(t),Q_2(t),\dots,Q_K(t)]}$.
 Then, a quadratic Lyapunov  function $L(\bold{Q}(t))$  is defined by \cite[Ch. 3]{neely2010stochastic} 
 \begin{align}\label{consta4}
L(\bold{Q}(t))=\dfrac{1}{2}\textstyle\sum_{k\in\mathcal{K}}Q^2_k(t).
 \end{align}
 The Lyapunov function measures the network congestion:  if the Lyapunov function is small, then all the queues are small, and if the Lyapunov function is large, then at least one queue is large. Therefore, by minimizing the expected change of the Lyapunov function from one slot to the next slot, queues $\{Q_k(t)\}_{k\in \mathcal{K}}$  can be stabilized \cite[Ch. 4]{neely2010stochastic}.
 
 The expected of the Lyapunov function from one slot to the next slot is defined as the drift in  the Lyapunov function, which is defined as 
  \begin{align}\label{consta5}
 \alpha(\bold{S}(t))=\mathbb{E}\left[L\left(\bold{Q}(t+1)\right)-L\left(\bold{Q}(t)\right)|\bold{S}(t)\right].
 \end{align}
According to the drift-plus-penalty minimization method, a control policy that minimizes the objective function of the optimization problem \eqref{eqo1} is obtained by minimizing the drift-plus-penalty in each slot $t$ \cite[Ch. 3]{neely2010stochastic}, i.e.,
 \begin{align}\label{consta6}
\alpha(\bold{S}(t))+V\textstyle\sum_{k\in \mathcal{K}}\textstyle\sum_{n\in\mathcal{N}}\mathbb{E}[p_{k,n}(t)],
\end{align}
  subject to the following constraints
  \begin{subequations}\label{eqo2}
 	\begin{align}\label{eq8a2}
 	&\textstyle\sum_{k\in\mathcal{K}}\rho_{k,n}(t)\le 1, \forall n\in \mathcal{N}, \forall t\\&\label{eqo32}
 	\textstyle\sum_{n\in\mathcal{N}}p_{k,n}(t)\le P_k^{\text{max}},\,\,\, k\in \mathcal{K}, \forall t
 	\\&\label{eq1o52}
 R_{k}(t)=\eta b_{k}(t),  k\in \mathcal{K}, \forall t
 	\\&\label{eq1o502}
 	\rho_{k,n}(t)\in\{0,1\},  k\in \mathcal{K}, n\in \mathcal{N}, \forall t
 	\\&\label{eq1o512}
 	b_{k}(t)\in\{0,1\},  k\in \mathcal{K},  \forall t,
 	\end{align}
 \end{subequations}
where $V\ge0$ is a parameter that represents  how much we emphasize on the objective function (power minimization). Therefore, by varying $V$, a desired trade-off between the  sizes of the queue backlogs  and objective function can be obtained. 

Since 
minimizing the objective function \eqref{consta6} is intractable,  we  minimize an upper bound of \eqref{consta6} in each slot $t$ \cite[Ch. 3]{neely2010stochastic}.
  To find an upper bound for \eqref{consta6}, we use the following inequality in which  for any $\hat{A}\ge0$, $\tilde{A}\ge0$, and $\bar{A}\ge0$ we have \cite[Ch. 3]{neely2010stochastic}
 \begin{align}\label{wrf01}
 \left(\max\left[\hat{A}-\tilde{A},0\right]+\bar{A}\right)^2\le\hat{A}^2+\tilde{A}^2+\bar{A}^2+2\hat{A}(\bar{A}-\tilde{A}).
 \end{align}
 By applying \eqref{wrf01} to \eqref{consta2}, an upper bound  for $ Q^2_k(t+1)$  is given as 
 \begin{align}\nonumber
& Q^2_k(t+1)\le Q^2_k(t)+\left(\Delta^{\text{max}}_k-\dfrac{1}{2}\right)^2+\delta^2_k(t+1)+2Q_k(t)\\&\label{021mb4} \left(\delta_k(t+1)-(\Delta^{\text{max}}_k-\dfrac{1}{2})\right).
 \end{align}
 By applying \eqref{021mb4} to \eqref{consta6}, an upper bound for \eqref{consta6} is given as
  \begin{align}\nonumber
  &\alpha(\bold{S}(t))+V\sum_{k\in \mathcal{K}}\sum_{n\in\mathcal{N}}\mathbb{E}[p_{k,n}(t)]\le V\sum_{k\in \mathcal{K}}\sum_{n\in\mathcal{N}}\mathbb{E}[p_{k,n}(t)]+\\&\nonumber\dfrac{1}{2}\mathbb{E}\bigg[\sum_{k\in\mathcal{K}}\!\bigg((\Delta^{\text{max}}_k-\dfrac{1}{2})^2+\delta^2_k(t+1)+2Q_k(t)\big(\delta_k(t+1)-\\&\nonumber(\Delta^{\text{max}}_k-\dfrac{1}{2})\big)\bigg)\bigg|\bold{S}(t)\bigg]=V\textstyle\sum_{k\in \mathcal{K}}\textstyle\sum_{n\in\mathcal{N}}\mathbb{E}[p_{k,n}(t)]\\&\nonumber+\dfrac{1}{2}\sum_{k\in\mathcal{K}}\bigg((\Delta^{\text{max}}_k-\dfrac{1}{2})^2+\mathbb{E}[\delta^2_k(t+1)|\bold{S}(t)]+2Q_k(t)\\&\label{021mb40}\big(\mathbb{E}[\delta_k(t+1)|\bold{S}(t)]-(\Delta^{\text{max}}_k-\dfrac{1}{2})\big)\bigg).
   \end{align}
   To derive the upper bound for \eqref{consta6}, we need to determine  $\mathbb{E}[\delta_k(t+1)|\bold{S}(t)]$ and $\mathbb{E}[\delta^2_k(t+1)|\bold{S}(t)]$. To this end, by using the  evolution of the AoI in \eqref{AoI1},  $\delta_k(t+1)$ and $\delta^2_k(t+1)$ are calculated as 
     \begin{align}\label{021mb401}
     &\delta_k(t+1)=b_k(t)+\left(1-b_k(t)\right)(\delta_k(t)+1),  k\in \mathcal{K} \\&\nonumber
     \delta^2_k(t+1)=b_k(t)+\left(1-b_k(t)\right)(\delta_k(t)+1)^2,  k\in \mathcal{K}.
        \end{align}
        By using the expressions in \eqref{021mb401}, $\mathbb{E}[\delta_k(t+1)|\bold{S}(t)]$ and  $\mathbb{E}[\delta^2_k(t+1)|\bold{S}(t)]$ are given as 
        \begin{align}\label{021mb4010}
        &\mathbb{E}[\delta_k(t+1)|\bold{S}(t)]\!=\!\mathbb{E}[b_k(t)]\!+\!(1\!-\!\mathbb{E}[b_k(t)])(\delta_k(t)\!+\!1),  k\in \mathcal{K} \\&\nonumber
                \mathbb{E}[\delta^2_k(t\!+\!1)|\bold{S}(t)]\!=\!\mathbb{E}[b_k(t)]\!+\!(1\!-\!\mathbb{E}[b_k(t)])(\delta_k(t)\!+\!1)^2, k\!\in\! \mathcal{K} .
        \end{align}
        
        By substituting  \eqref{021mb4010} into the right hand side of  \eqref{021mb40}, the upper bound for \eqref{consta6} is given as 
        \begin{align}\nonumber
        &V\sum_{k\in \mathcal{K}}\sum_{n\in\mathcal{N}}\mathbb{E}[p_{k,n}(t)]+\dfrac{1}{2}\sum_{k\in\mathcal{K}}\bigg( (\Delta^{\text{max}}_k-\dfrac{1}{2})^2+(\delta_k(t)+1)^2\\&\nonumber+(2Q_k(t)-1)(\delta_k(t)+1)+ \mathbb{E}[b_k(t)]\big(1-(\delta_k(t)+1)^2\\&\label{uppervn} -{{2Q_k(t)}}\delta_k(t)\big)\bigg).
        \end{align}
        
        Next, we explain the proposed dynamic algorithm to  solve the optimization problem \eqref{eqo1}. The main steps of the algorithm are summarized in  Algorithm 1.  
        The algorithm observes the channel states $\{h_{k,n}(t)\}_{k\in \mathcal{K},n\in \mathcal{N}}$ and network state  $\bold{S}(t)$ in each time slot $t$ and makes a control action to minimize \eqref{uppervn} subject to the constraints \eqref{eq8a2}-\eqref{eq1o512}. Note that the drift-plus penalty method exploits the \textit{opportunistically minimize an expectation} \cite[Ch. 8]{neely2010stochastic}  to solve the subproblem in each slot.  To solve the optimization problem \eqref{eqoi1} in each slot, we confine to use the   exhaustive search algorithm.
        
          \begin{algorithm}[t]
        	{   \caption{Proposed solution algorithm for problem \eqref{eqo1}}
        		\label{table-1}
        		Step 1: \textbf{initialization}: set ${t=0}$, set $V$, and  initialize $~~~~~~~~~~~~~$$\{Q_k(0),\delta_k(0)\}_{k\in \mathcal{K}}$,
        		\\
        		\textbf{for}  each time slot $t$ \textbf{do}
        		\\
        		Step 2: Sampling action, transmit power, and subchannel $~~~~~~~~~$assignment: obtain $\{p_{k,n}(t),\rho_{k,n}(t)\}_{k \in\mathcal{K}, n \in\mathcal{N}}$ and $~~~~~~~~~$$\{b_{k}(t)\}_{k \in\mathcal{K}}$ by solving the following optimization $~~~~~~~~~~~~~$problem: 
        		\begin{align}\label{eqoi1}
        		&\text{minimize} \,\,V\sum_{k\in \mathcal{K}}\sum_{n\in\mathcal{N}}p_{k,n}(t)+\\&\nonumber\hspace{1.5cm}\dfrac{1}{2}\sum_{k\in\mathcal{K}}\bigg( 
        		b_k(t)\big(1-(\delta_k(t)+1)^2-2Q_k(t)\delta_k(t)\big)\bigg)\\&\nonumber
        		\text{subject\,\,to}\hspace{.2cm}
        		\eqref{eq8a2}-\eqref{eq1o512},
        		\end{align}
        		$~~~$with variables $\{p_{k,n}(t),\rho_{k,n}(t)\}_{k \in\mathcal{K}, n \in\mathcal{N}}$ and $~~~~~~~~~~~$$\{b_{k}(t)\}_{k \in\mathcal{K}}$,
        		\\
        		Step 3: Queue update:  update $\{Q_k(t+1),\delta_k(t+1)\}_{k\in \mathcal{K}}$ $~~~~~~~~~~~~$using \eqref{consta2} and  \eqref{021mb401},
        		\\
        		$~~~~~~~~~~~~$Set $t=t+1$, and go to Step 2,
        		\\
        	\textbf{end for}
        	}
        \end{algorithm}
        \section{Numerical Results}\label{Numerical Results}
        In this section, we evaluate the performance of the proposed dynamic control algorithm presented in Algorithm 1. Due to the complexity of the exhaustive search solution used to solve the optimization problem \eqref{eqoi1}, we evaluate the performance of the system with a small number of sensors and subchannels. We consider ${K=2}$ sensors placed in a two-dimensional plane and ${N=2}$ subchannels with bandwidth ${W=180}$ kHz. The  coordinate of sensor $1$ is $(0,300)$, the coordinate of sensor $2$ is  $(300,0)$, and the coordinate of the sink is $(0,0)$.  The channel coefficient from sensor $k$ to the sink over subchannel $n$ at slot $t$ is modeled by $h_{k,n}(t)=(d_k/d_0)^\xi c_{k,n}(t)$, where $d_k$ is the distance from sensor $k$  to the sink, $ d_0 $ is the
        far field reference distance,  $\xi$ is the path loss exponent,
        and $c_{k,n}(t)$ is a Rayleigh distributed random coefficient.   Accordingly,  $(d_k/d_0)^\xi$  represents  large
        scale fading and the term ${c_{k,n}(t)}$ denotes small scale Rayleigh fading. We set ${\xi=-3}$, $ {d_0=1} $, maximum acceptable average AoI of sensors ${\Delta_{k}^{\text{max}}=4, \forall k}$, the size of each  packet  ${\eta=600}$ Bytes, and the parameter of Rayleigh distribution is $0.5$. 

\begin{figure}
	\centering
	\includegraphics[scale=0.4]{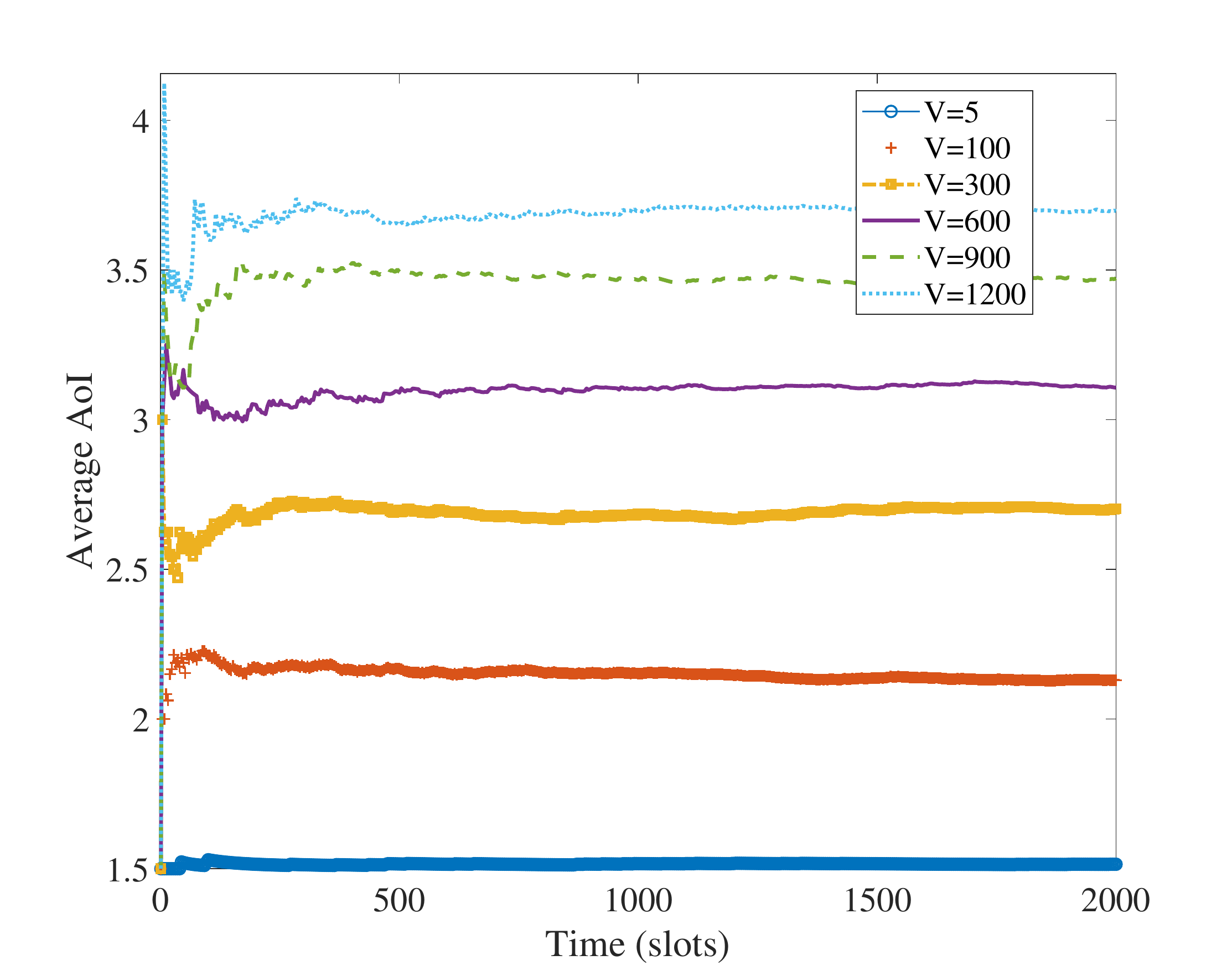}
	\caption{ Average AoI of sensor 1 as a function of $V$. }
		 		\vspace{-5mm}
	\label{f02}
  \end{figure}

        Fig. \ref{f02} depicts the average AoI of sensor 1 as a function of $V$.  According to this figure, when  $V$ increases, the average AoI of sensor 1 increases as well. This is because  when $V$ increases, the backlogs of the virtual queues associated to the time average AoI constraints \eqref{eqo5}  increase.  We can also observe that the average AoI of the sensor is always smaller than   the maximum acceptable average AoI ${\Delta_{k}^{\text{max}}=4}$.
        
\begin{figure}
	\centering
	\includegraphics[scale=0.41]{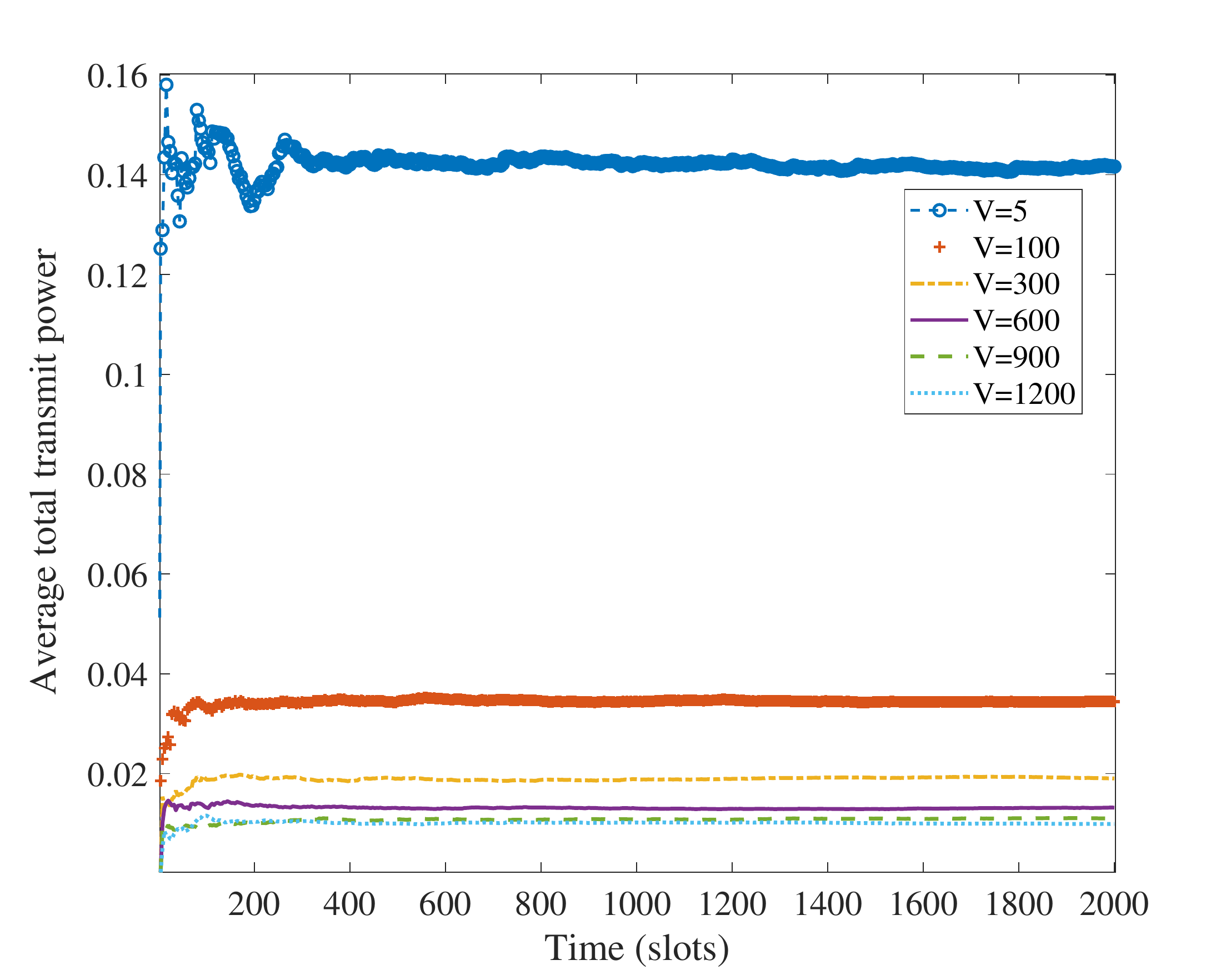}
	\caption{Average total transmit power of the sensors as a function of $V$. }
		 		\vspace{-5mm}
	\label{f03}
  \end{figure}

        Fig. \ref{f03} illustrates the time average total transmit power as a function of $V$. The figure shows that when  $V$ increases, the average total transmit power decreases. This is because  when $V$ increases, more emphasis is set to minimize the total transmit power in the objective function of optimization problem \eqref{eqoi1}.
        
\begin{figure}
	\centering
	\includegraphics[scale=0.39]{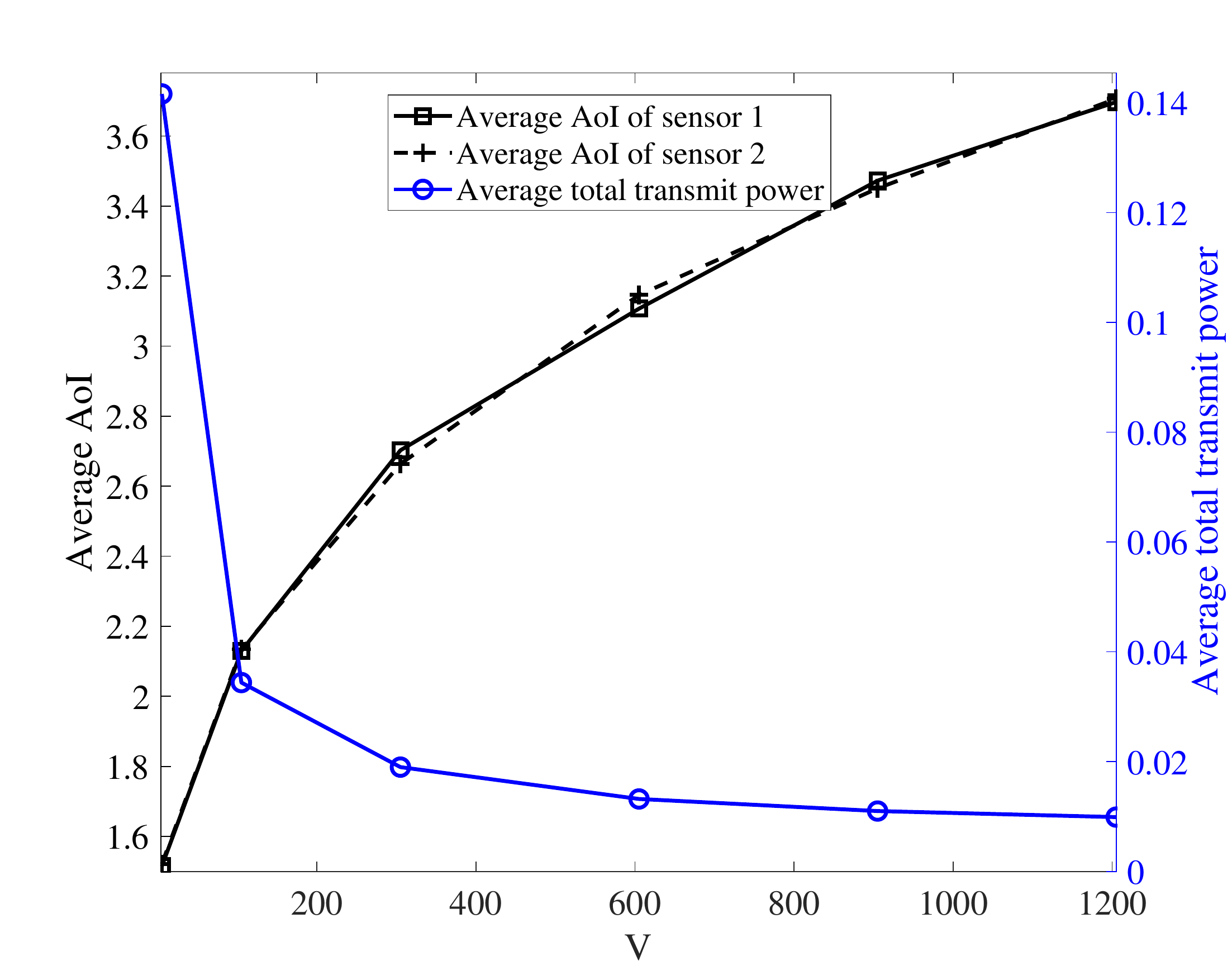}
	\caption{Trade-off between the average total transmit power and average AoI of sensor 1 and sensor 2 as a function of $V$. }
		 		\vspace{-5mm}
	\label{f01}
\end{figure}

        Fig. \ref{f01} illustrates the trade-off between the average AoI  and  average total transmit power of the sensors for different values of $V$. By increasing  $V$ the average AoI of different sensors increases and the average total transmit power decreases. 
        Note that the average AoI of sensors remains always smaller than the  maximum acceptable average AoI.
        
        From   Figs. \ref{f02} and \ref{f01}, we observe that by increasing $V$ sufficiently high, the AoI values of the sensors eventually reach the maximum acceptable average AoI. Similarly, as it can be seen in  Figs. \ref{f03} and \ref{f01},  for the high values of $V$,  the average total transmit power of the sensors starts to saturate into a certain level.

%
%
%
%


\section{Conclusions}\label{Conclusion}
In this paper, we considered a status update system consisting of a set of sensors that can control the sampling action. The status update packets of the sensors are transmitted by sharing a set of orthogonal subchannels  in each slot.  We formulated an optimization problem to minimize the  time average total transmit power of sensors with time average AoI and maximum power constraints for each sensor. To solve the proposed optimization problem, we used the Lyapunov drift-plus-penalty method. This method provides a trade-off between the average total transmit power and the average AoI of the sensors  which were  shown in the  numerical experiments.

   \section*{Acknowledgements}
  This research has been financially supported by the Infotech Oulu, the Academy of Finland (grant 323698), and Academy of Finland 6Genesis Flagship (grant 318927). M. Codreanu would like to acknowledge the support of the European Union's Horizon 2020 research and innovation programme under the Marie Sk\l{}odowska-Curie Grant Agreement No. 793402 (COMPRESS NETS). M. Moltafet would like to acknowledge the support of Finnish Foundation for Technology Promotion.
\bibliographystyle{IEEEtran}
\bibliography{Stochastic-Optimization}

\end{document}